\documentclass[final,1p,times]{elsarticle} % do not change
\usepackage{graphicx} % do not change
\usepackage{amssymb} % do not change
\usepackage{amsthm} % do not change
\usepackage{lineno} % do not change

\journal{Nuclear Physics A} % do not change
\begin{document} % do not change

\begin{frontmatter} % do not change

%% QM09Author: please enter your  
%% Title, author and address info here; please do not use footnotes

% Your Title - please modify
  \title{Probing medium-induced energy loss with direct jet
    reconstruction in p+p and Cu+Cu collisions at PHENIX}

% Principle author, and co-authors - please modify
\author{Yue-Shi Lai$^{a}$ for the PHENIX collaboration}

% Address - please modify
% note that if you have authors from several institutions, we recommend
% labelling these [a], [b], [c] etc.
\address[a]{Columbia University, % label [a]
538 West 120th Street,
New York, NY, 10027-7061, USA, and
Nevis Laboratories,
136 South Broadway,
Irvington, NY 10533-2508, USA}

\begin{abstract} % do not change
%% Text of abstract goes here - please modify
  We present the application of a new jet reconstruction algorithm
  that uses a Gaussian filter to locate and reconstruct the jet energy
  to $p + p$ and heavy ion data from the PHENIX detector. This
  algorithm is combined with a fake jet rejection scheme that provides
  efficient jet reconstruction with an acceptable fake rate. We show
  our first results on the measured jet spectra, and on jet--jet
  angular correlation in p+p and Cu+Cu collisions.
\end{abstract} % do not change

\end{frontmatter} % do not change

%% QM09: we keep linenumbers at least for initial version
%\linenumbers % do not change

The study of jet physics at RHIC plays an important role in
understanding the parton--medium interaction, in-medium fragmentation
properties, and applied to the RHIC spin program, also the study of
the proton structure. We therefore set out to systematically
investigate the feasibility of, and the appropriate approach of
performing jet reconstruction using a limited aperture detector like
the PHENIX central arms~\cite{Adcox:2003zm} and including heavy ion
collision systems. Neither is a well-studied aspect of jet
reconstruction, and both have been considered as
challenging~\cite{Adler:2005ad}.

We observed that the flat weighting in traditional jet reconstruction
algorithms is particularly prone to fluctuations at large angle. A
similiar effect also exists at the PHENIX central arm edges, here due
to the lack of balancing fragments. Both issues can be effectively
addressed by a nonflat weighting that smoothly dampens large angle
fragments. Furthermore, the energy flow variable as proposed by
Sterman et al.~\cite{Berger:2003iw} suggests that angular convolution
of the event $p_T$ with a continuous distribution can provide an
effective description of QCD processes. All these reasions pointed us
towards a seedless jet definition as the output of a filter, e.g. by
using the Gaussian weighting
\begin{equation}
  p_T^\mathrm{filt}(\eta, \phi) =
  \int_{\eta_0}^{\eta_1}\int_{-\pi}^{\pi} d\eta'\, d\phi'\, p_T(\eta',
  \phi') e^{-((\eta - \eta')^2 + (\phi - \phi')^2)/2\sigma}
\end{equation}
and reconstructing the resulting local maxima in $p_T^\mathrm{filt}$
as jets. The Gaussian filter kernel has the property of no maximum
creation, and the jet definition therefore becomes trivially collinear
and infrared safe. A description of its efficient implementation and
comparison of the behavior against other jet definitions for p+p
collisions is given in \cite{Lai:2008zp}.

Historically, there have been similar attempts to define jets using a
filter. The British-French-Scandinavian collaboration used a jet
definition via convolution~\cite{Albrow:1979yc}, which notably
involves a Gaussian kernel with $\sigma = 0.5$ and predates the
Snowmass accord on the cone algorithm~\cite{Huth:1990mi}. However,
little subsequent study and application of the Gaussian filter was
made.

\begin{figure}[ht]
\centering
\includegraphics[width=0.75\textwidth]{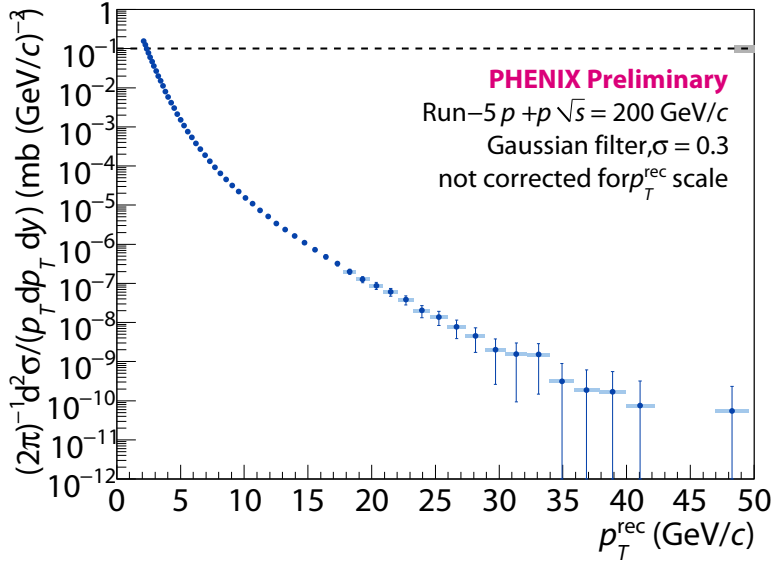}
\caption[]{PHENIX Run-5 p+p raw jet spectrum using $\sigma = 0.3$
  Gaussian filter}
\label{run_5_p_p_spectrum}
\end{figure}

\begin{figure}[p]
\centering
\includegraphics[width=0.75\textwidth]{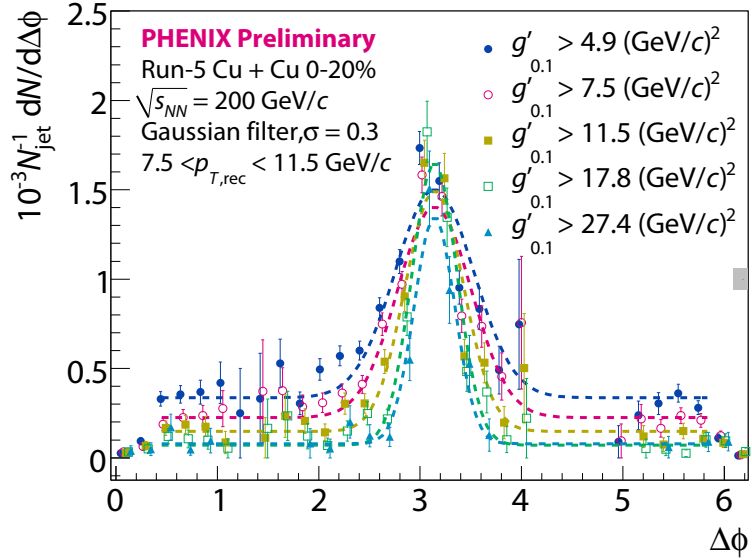}
\caption[]{Run-5 Cu+Cu $\Delta\phi$ distribution for symmetric dijets
  with $7.5\,\mathrm{GeV}/c < p_T^\mathrm{rec} < 11.5\,\mathrm{GeV}/c$
  and different fake rejection levels}
\label{run_5_cu_cu_dphi_fr}
\end{figure}

\begin{figure}[p]
\centering
\includegraphics[width=0.75\textwidth]{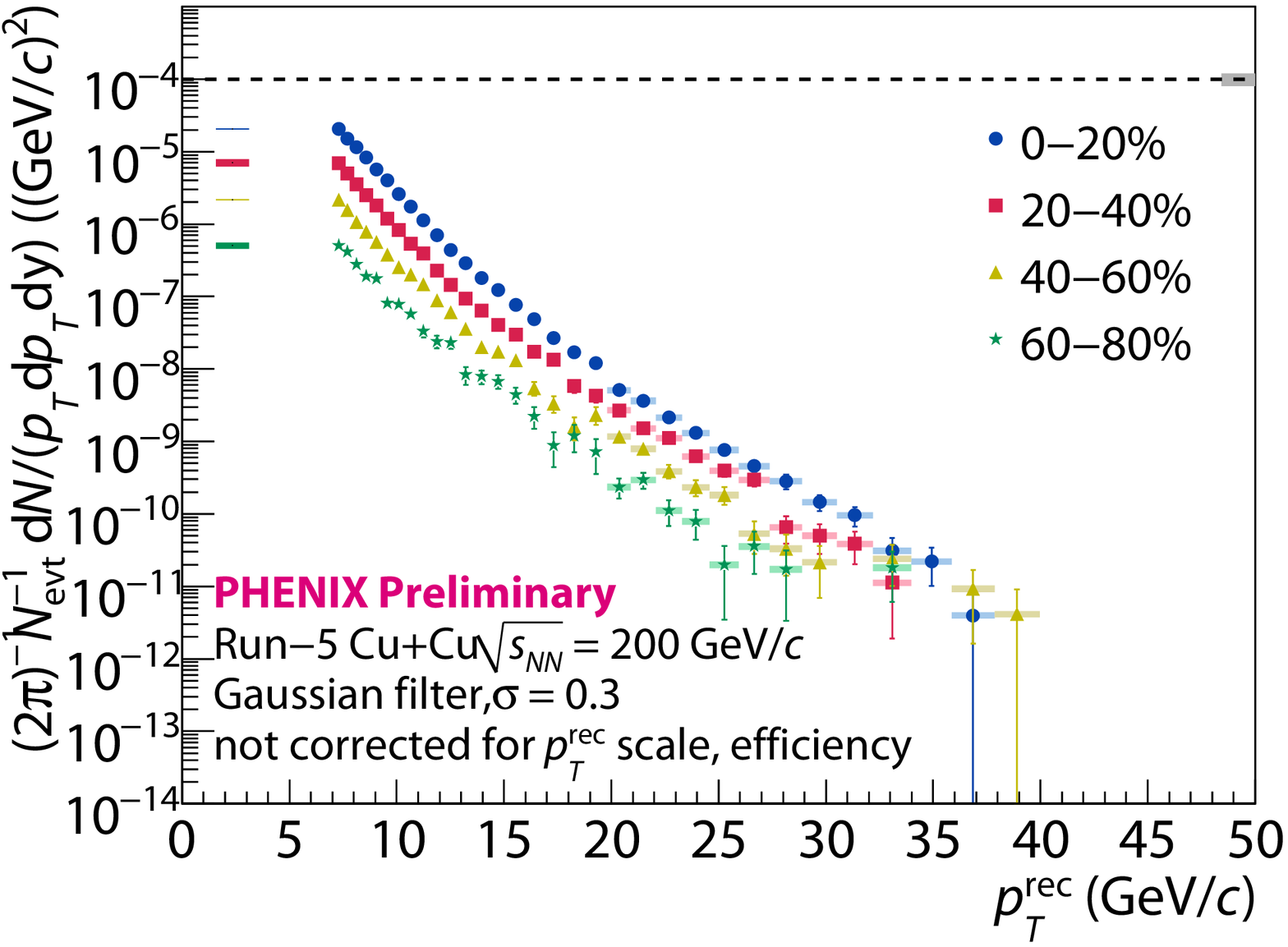}
\caption[]{PHENIX Run-5 fake rejected Cu+Cu raw jet spectrum using
  $\sigma = 0.3$ Gaussian filter, $g_{\sigma_\mathrm{dis}} >
  17.8\,(\mathrm{GeV}/c)^2$ fake rejection}
\label{run_5_cu_cu_spectrum}
\end{figure}

A direct application of jet reconstruction algorithms to heavy ion
collisions is known to give rise to false apparent jet production
rates~\cite{Stewart:1990wa}. Unlike at LHC, the collision energy at
RHIC results in a significantly lower jet production rate, e.g. for
Cu+Cu at $\sqrt{s_{NN}} = 200\,\mathrm{GeV}$ and $p_T =
10\,\mathrm{GeV}$, $(2\pi)^{-1} N_\mathrm{evt}^{-1} dN/(p_T dp_T dy)
\approx 10^{-6} (\mathrm{GeV}/c)^{-2}$ (cf.\ Fig.\
\ref{run_5_cu_cu_spectrum}. Jet production at these rates is easily
submerged under background fluctuations. The direct fake rejection is
an effective techique to filter out fake jets, since it provides
jet-by-jet information about its likelihood to be the result of a real
hard scattering.

This leads us to develop a fake rejection strategy that can achieve a
stronger level of fake jet suppression than what has been proposed for
the LHC (e.g.\ \cite{Grau:2008ed}), while preserving a fast rise of
efficiency to unity. We define a fake rejection discriminant
\begin{equation}
  g_{\sigma_\mathrm{dis}}(\eta, \phi) = \sum_{i \in \mathrm{fragment}}
  p_{T,i}^2 e^{-((\eta_i - \eta)^2 + (\phi_i -
    \phi)^2)/2\sigma_\mathrm{dis}},
\end{equation}
where $(\eta, \phi)$ is approximately the jet axis, but we allow it to
shift, since jets may not be perfectly centered due to background
fluctuation. The discriminant size is chosen to be
$\sigma_\mathrm{dis} = 0.1$.

\begin{figure}[ht]
\centering
\includegraphics[width=0.75\textwidth]{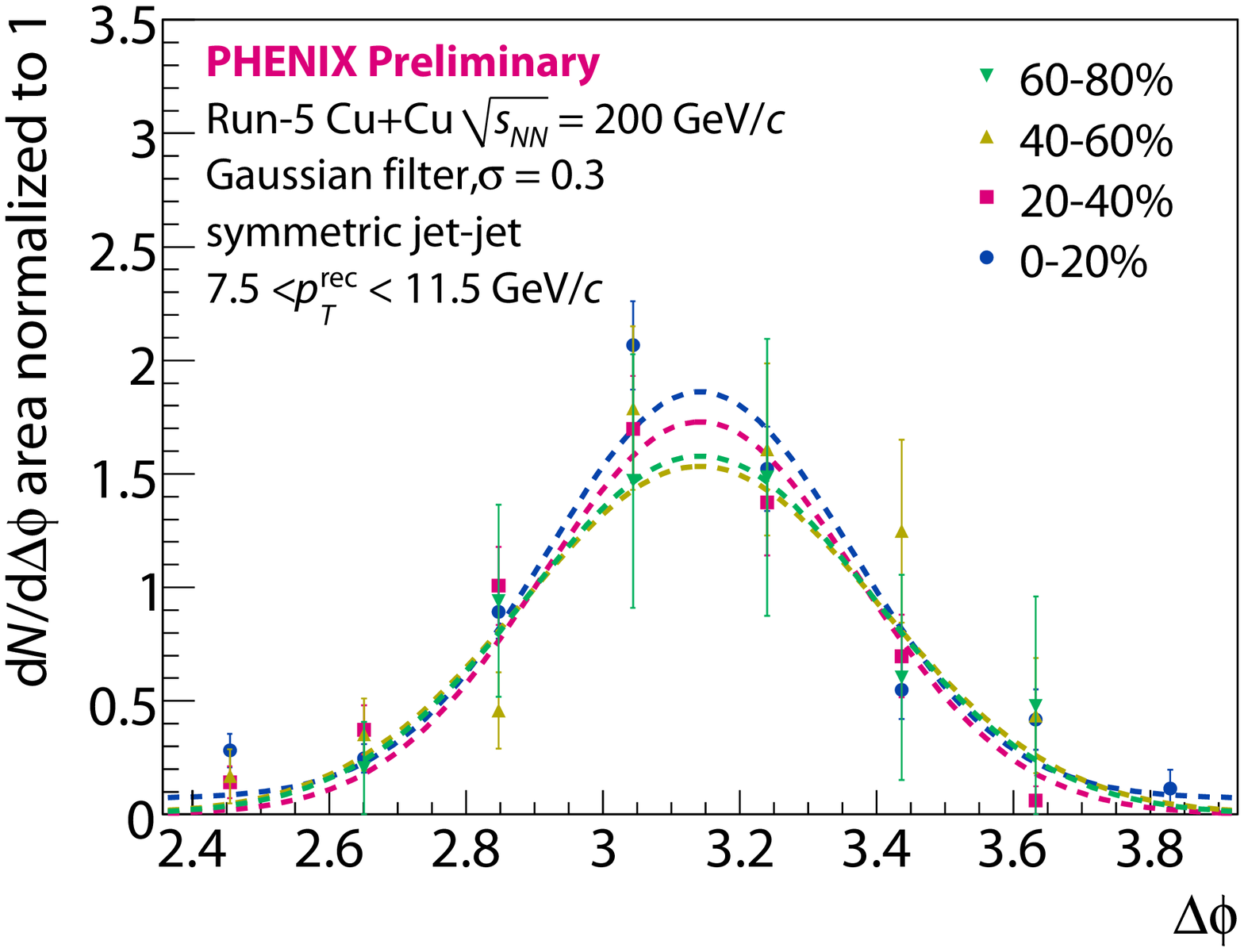}
\caption[]{Run-5 Cu+Cu $\Delta\phi$ distribution for symmetric dijets
  with $7.5\,\mathrm{GeV}/c < p_T^\mathrm{rec} <
  11.5\,\mathrm{GeV}/c$, $g_{\sigma_\mathrm{dis}} >
  17.8\,(\mathrm{GeV}/c)^2$ fake rejection, and different
  centralities}
\label{run_5_cu_cu_dphi_zoomed}
\end{figure}

Fig.\ \ref{run_5_p_p_spectrum} shows the reconstructed p+p jet
spectrum. Subsequent unfolding using the \textsc{geant} derived
detector response shows that the spectrum is consistent with
\cite{Abelev:2006uq}. The loss of $n$ and $K^0_L$ shifts the
reconstructed energy scale $p_T^\mathrm{rec}$ below the true energy
scale. The fact that the $p_T$ reaches out to $40\,\mathrm{GeV}/c$
demonstrates that for the study of jet physics, the fast event readout
of PHENIX easily offsets its small aperture.

Fig.\ \ref{run_5_cu_cu_dphi_fr} shows the reconstructed Cu+Cu dijet
$\Delta\phi$ distribution for different fake rejection thresholds and
symmetric dijets with $7.5\,\mathrm{GeV}/c < p_T^\mathrm{rec} <
11.5\,\mathrm{GeV}/c$. The effect of the fake rejection to suppress
fake jets that has a random orientation against jets produced by hard
scattering is clearly visible. For $g_{\sigma_\mathrm{dis}} >
4.9\,(\mathrm{GeV}/c)^2$, the pedestal translates into a fake rate
about 10 times the jet yield. The effect of the fake rejection
saturates at approximately $g_{\sigma_\mathrm{dis}} >
17.8\,(\mathrm{GeV}/c)^2$, where the fake contribution is constrained
to below $10\%$.

Fig.\ \ref{run_5_cu_cu_spectrum} shows the reconstructed Cu+Cu jet
spectrum for different centralities and $g_{\sigma_\mathrm{dis}} >
17.8\,(\mathrm{GeV}/c)^2$. After the application of fake rejection, a
consistent power-law shape across all centralities is evident.

Fig.\ \ref{run_5_cu_cu_dphi_zoomed} shows the reconstructed Cu+Cu
dijet $\Delta\phi$ distribution with $g_{\sigma_\mathrm{dis}} >
17.8\,(\mathrm{GeV}/c)^2$ and for symmetric dijets with
$7.5\,\mathrm{GeV}/c < p_T^\mathrm{rec} < 11.5\,\mathrm{GeV}/c$ in
different centralities. Also here, the application of fake rejection
results in a $\Delta\phi$ distribution that is consistent across all
centralities (within statistical errors).

In summary, we showed that jet physics can be effectively studied
using PHENIX and the Gaussian filter-based jet reconstruction
algorithm we proposed. The proper rejection of fake jets is an
important aspect of applying jet reconstruction to heavy ion
collisions. We have shown our first results in p+p and Cu+Cu
collisions. Further studies, including the measurement of $R_{AA}$ and
the fragmentation function, are underway.

 % do not change 

\begin{thebibliography}{00} % do not change 

\bibitem{Adcox:2003zm} K. Adcox et al., {\it Nucl. Instrum. Meth.}
  {\bf A499} (2003) 469--479.

\bibitem{Adler:2005ad} S.S. Adler et al., {\it Phys. Rev.} {\bf C73}
  (2006) 054903.

\bibitem{Berger:2003iw} C.F. Berger, T. K{\'u}cs, and G. Sterman, {\it
    Phys. Rev.} {\bf D68} (2003) 014012.

\bibitem{Lai:2008zp} Y.-s. Lai, B.A. Cole, [arXiv:0806.1499]

\bibitem{Albrow:1979yc} M. G. Albrow et al., {\it Nucl. Phys.} {\bf
    B160}, 1 (1979).

\bibitem{Huth:1990mi} J.E. Huth et al., in {\it Research directions
    for the decade: proceedings of the 1990 Summer Study on High
    Energy Physics} (1990) 134--136.

\bibitem{Stewart:1990wa} C. Stewart et al., {\it Phys. Rev.} {\bf D42}
  (1990) 1385--1395.

\bibitem{Grau:2008ed} N. Grau et al. [ATLAS collaboration],
  [arXiv:0810.1219]

\bibitem{Abelev:2006uq} B.I. Abelev et al., {\it Phys. Rev. Lett.}
  {\bf 97} (2006) 252001.


\end{thebibliography}
\end{document}